\documentclass[conference]{IEEEtran}
\IEEEoverridecommandlockouts
\usepackage{cite}
\usepackage{amsmath,amssymb,amsfonts}
\usepackage{algorithmic}
\usepackage{graphicx}
\usepackage{textcomp}
\usepackage{xcolor}
\usepackage{tikz}
\usepackage{flushend}
\usepackage[left=0.62in, right = 0.62in, top=0.7in, bottom = 0.95in]{geometry}
\def\BibTeX{{\rm B\kern-.05em{\sc i\kern-.025em b}\kern-.08em
    T\kern-.1667em\lower.7ex\hbox{E}\kern-.125emX}}
\begin{document}

\title{A SKG Security Challenge: Indoor SKG Under an On-The-Shoulder Eavesdropping Attack\\
\thanks{This work is financed on the basis of the budget passed by the Saxon State Parliament, and partially funded by Hexa-X-II project, receiving funding from the Smart Networks and Services Joint Undertaking (SNS JU) under the European Union’s Horizon Europe research and innovation programme under Grant Agreement No 101095759.}
}

\author{
    \IEEEauthorblockN{Amitha Mayya$^1$, Miroslav Mitev$^1$, Arsenia Chorti$^{1,2}$, Gerhard Fettweis$^1$}
    \IEEEauthorblockA{$^1$Barkhausen Institut, Dresden, Germany \\
    $^2$ETIS UMR 8051, CYU, ENSEA, CNRS, Cergy, France\\
        \{amitha.mayya, miroslav.mitev, gerhard.fettweis\}@barkhauseninstitut.org, {arsenia.chorti@ensea.fr}
        }}

\maketitle

\begin{abstract}
Physical layer security (PLS) is seen as the means to enhance physical layer trustworthiness in 6G. This work provides a proof-of-concept for one of the most mature PLS technologies, i.e., secret key generation (SKG) from wireless fading coefficients during the channel's coherence time. As opposed to other works, where only specific parts of the protocol are typically investigated, here, we implement the full SKG chain in four indoor experimental campaigns. In detail, we consider two legitimate nodes, who use the wireless channel to extract secret keys and a malicious node placed in the immediate vicinity of one of them, who acts as a passive eavesdropper. To estimate the final SKG rate we evaluate the conditional min-entropy by taking into account all information available at the eavesdropper. Finally, we use this paper to announce the first ever physical layer security challenge, mirroring practices in cryptography. We call the community to scrutinize the presented results and try to ``break'' our SKG implementation. To this end, we provide, i) the full dataset observed by the eavesdroppers, ii) $20$ blocks of $16-$byte long ciphertexts, encrypted using one-time pad with $20$ distilled secret keys, and, iii) all codes and software used in our SKG implementation. An attack will be considered successful if any part(s) of the plaintext are successfully retrieved.

\end{abstract}

\begin{IEEEkeywords}
Security challenge, secret key generation, physical layer security, 6G, experimental campaign.
\end{IEEEkeywords}

\section{Introduction}
In the  sixth-generation (6G) of wireless, due to the increased complexity of interconnected heterogeneous systems, the need to ensure trustworthy communications becomes very important. 
To ensure security, PLS exploits the physical properties of the wireless channel and device hardware as sources of randomness~\cite{Chorti_book, PLS_Chorti}. A promising approach that can be used in hybrid crypto-PLS systems to achieve confidentiality at the physical layer is the PLS-based secret key generation (SKG). SKG makes use of the reciprocity and randomness of the wireless channel as a source of entropy~\cite{mitev2020authenticated, PLS_review}. While different studies have focused on specific parts of the protocol, there are only few that have implemented the full SKG chain~\cite{ZHANG}.

A typical assumption in the SKG literature, based on Jake's model, is that the channel decorrelates at a distance of half-wavelength  \cite{Goldsmith},\cite{half_wavelength}. However, this assumption holds only when the environment has infinite uniformly distributed scattering~\cite{He2016linksignature}. Thus, in practice, it is important to account for the correlations between legitimate nodes' and eavesdropper's observations.To address this point, in this work, we focus exclusively in passive attacks. For active attacks please refer to our contributions in \cite{mitev_globecom_2019, mitev_2022_access, thuy_letter, chorti_ICC, srinivasan2021use}. 

In this study, the shared random component of the reciprocal channel is extracted using frequency modulation continuous waveform (FMCW) signals. Power observations at different frequency subbands are obtained by passing the received observations through a filterbank~\cite{Zoli_EURASIP}. 
In our earlier works~\cite{MITEV_vtc2022, MITEV_globecom2022}, we evaluated the filterbank approach analytically and through simulations. Instead, in this work we obtain measurements in real-life setups and provide practical validation.

The power observations at the output of the filterbank are converted into information bits using a library of quantizers.
Due to channel noise and imperfect estimation, the observations at each party could differ. This is corrected during the information reconciliation step using distributed source coding techniques \cite{Mahdi}.
Finally, during privacy amplification, potential information leakage is accounted for in the hashing rate. The leakage is measured using a conditional min-entropy estimator. Particularly, we use the fast black-box leakage estimation (F-BLEAU)~\cite{FBLEAU} which gives an estimation on the number of unpredictable and random bits. When this estimator is used to evaluate min-entropy, it is equivalent to the \textit{most common value} estimate of the NIST \cite{NIST_min_entropy} suite. Hashing is performed using SHA-256 which is considered to be a one-way collision resistant function. 


In order to command the confidence of academic, industrial and end users, we posit that PLS schemes should be subjected to the same level of scrutiny as in cryptography. Cryptographic primitives such as AES, DES, etc., have all been subjected to "security challenges" \cite{security_challenge}. With this paper, we would like to announce the first ever PLS security challenge \textit{aimed at validating the unbreakability of the secret keys generated through the SKG protocol performed on real indoor datasets.} We invite the readers to try and regenerate a set of the distilled secret keys given all the information available at the eavesdropper. The readers are also given the Python script for our SKG  implementation, including filterbank, quantization, reconciliation codes, the privacy amplification and the conditional min-entropy estimator. 



The rest of the paper is organized as follows. Section~\ref{sec:sys_model} describes the system model, Section~\ref{sec:SKG} presents the SKG protocol, Section~\ref{sec:measurement_campaign} describes the experimental setup scenarios, Section~\ref{sec:results} gives the detailed evaluation of each step of the SKG protocol, Section~\ref{sec:sec_challenge} introduces the security challenge, and Section~\ref{sec:conclusion} concludes this paper.

\section{System model} \label{sec:sys_model}
In this work, two legitimate users (Alice and Bob) generate  secret keys from the channel measurements. An eavesdropper (Eve) mounts a passive attack by intercepting their exchange. Alice and Bob send linear complex chirp signals using time division duplex (TDD) to obtain channel measurements. A chirp signal in baseband can be expressed as $x(t) =  \frac{1}{T}e^{j \pi c t^2}$, where $T$ denotes the duration of the chirp, $c=B/T$ is the chirp rate and $B$ denotes bandwidth.
The received signals are given as: 
\begin{equation}
     y_l(t)=x(t)*h_l(t)+w_l(t), \label{eq:A-measure}
\end{equation}
where $w_l(t)$ is a noise variable and $l\in{\{A,B,E\}}$ represents Alice, Bob and Eve respectively. The channel state information (CSI), $h_l$ between Alice and Bob is reciprocal when measured within the coherence time, i.e., $h_A(t) = h_B(t)$. Its correlation with Eve's CSI, $h_E(t)$, depends on the environment and Eve's location.

In our measurement campaign Eve is located very close to Bob (on-the-shoulder attack). The following section describes the SKG protocol through which Alice and Bob generate secret keys. 

\section{SKG Protocol} \label{sec:SKG}
\begin{figure}[!t]
\centering
\includegraphics[clip, trim=8.5cm 1.6cm 8.5cm 1.9cm,width=0.35\textwidth]{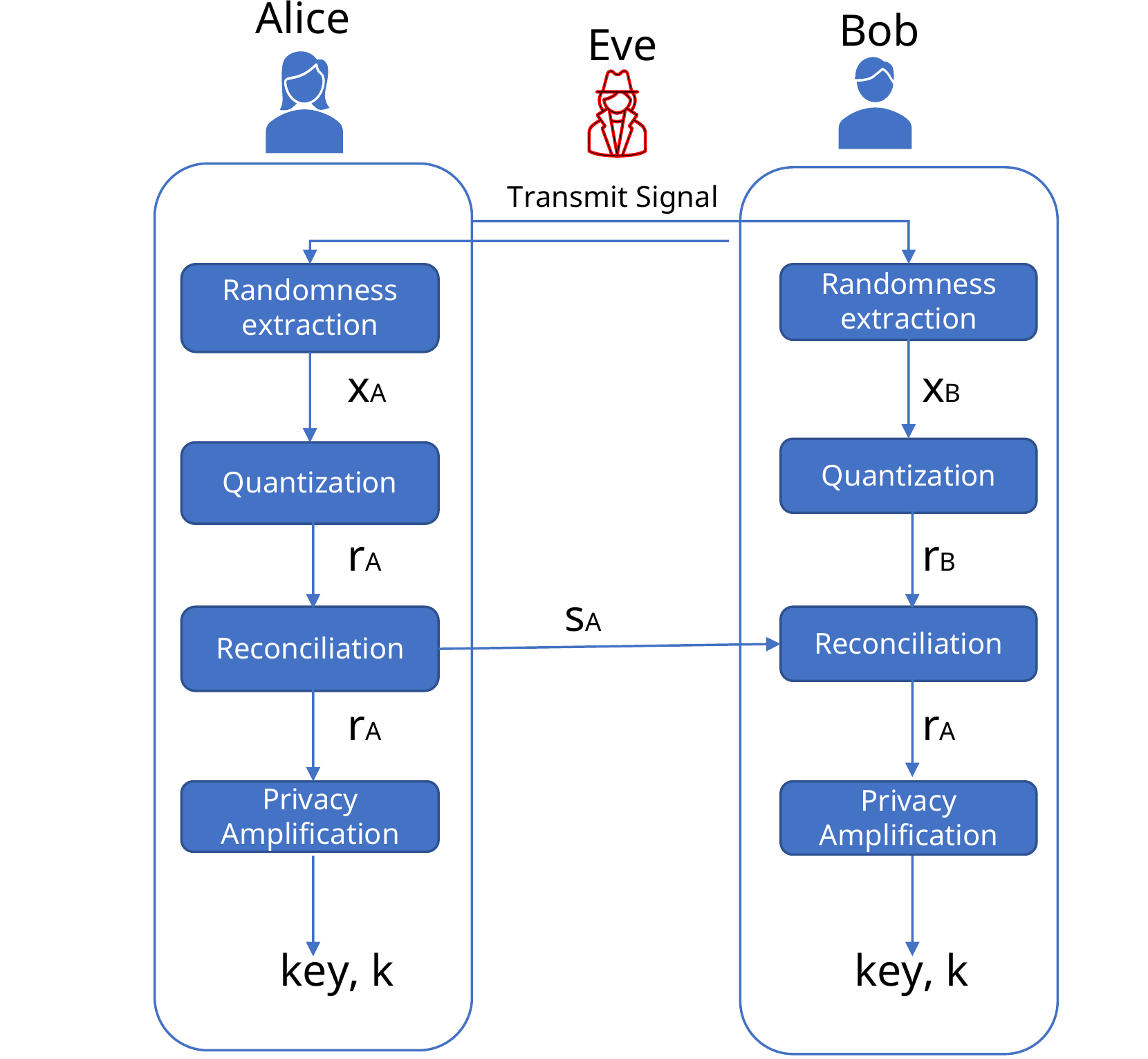}
\caption{Steps of the SKG protocol.}
\label{fig:SKGProtocol}
\vspace{-0.5cm}
\end{figure}

Fig.~\ref{fig:SKGProtocol} shows the SKG protocol followed in this work. In the advantage distillation phase the channel is measured, then quantized, so that a bit sequence is obtained. Bob's observation is reconciled to Alice's in the information reconciliation, while finally any potential leakage due to correlations with Eve's measurements is removed with privacy amplification. 

\subsection{Randomness extraction}
The received signals, $y_A, y_B, y_E$ are fed to a filterbank composed of $K$ filters. Each filter has bandwidth of $B/K$ and center frequency $f_k$ calculated as $f_k=-\frac{B(K-2k+1)}{2K}$. The impulse response of each individual filter $k = 1,..,K$ is
\begin{equation}
    g_k(t) = g(t)e^{j 2 \pi f_k(t)},
\end{equation}
where $g(t)$ is a protyping filter. Channel randomness is extracted from power measurements at the output of each filter. 
The obtained vector of power measurements at Alice, Bob and Eve are given as $\mathbf{{p}}_l=\left[{\hat{p}}_{{l,1}} \cdots {\hat{p}}_{{l,K}} \right]$ with $l\in{\{A,B,E\}}$. 

\begin{figure}[!t]
    \centering
    \includegraphics[clip, trim=1cm 0cm 3cm 0cm,width=0.5\textwidth]{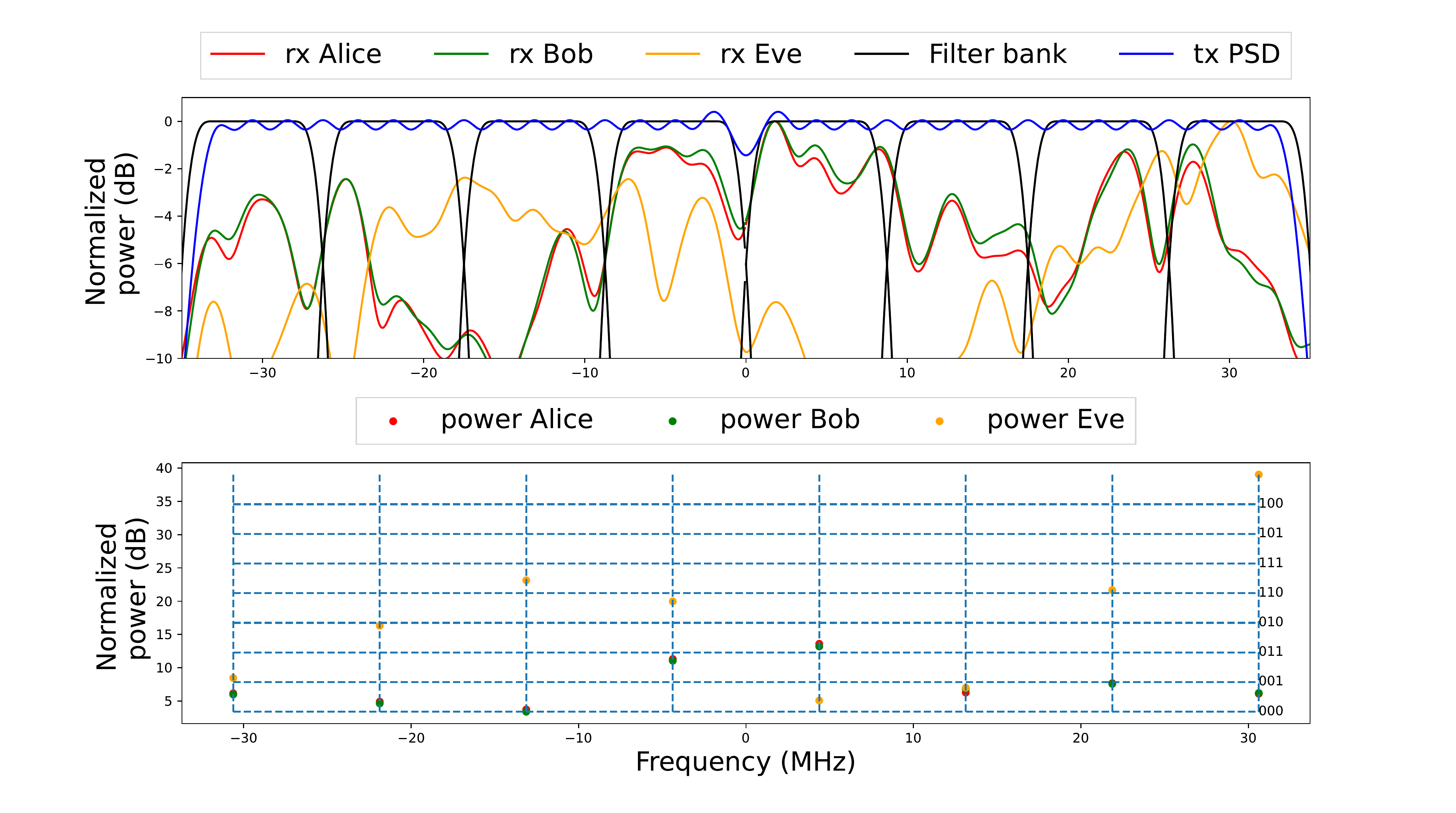}
    \caption{TOP: Power spectral density (PSD) of transmit signal and receive signals (Alice, Bob and Eve) and filterbank composed of $8$ filters. BOTTOM:power measurements at each subband aty Alice, Bob and Eve quantizer outputs represented in 3 bit Gray codes.}
    \label{fig:filterbank}
    \vspace{-0.4cm}
\end{figure}

Fig.~\ref{fig:filterbank} illustrates the power spectral density (PSD) of transmit signal and received signals at Alice, Bob and Eve measured using Welch’s method. The black curve represents the filterbank used to convolve the received signals. The time averaged power measurements are depicted by the dots for each frequency band. Channel correlations between Alice and Bob result in similar power measurement for each frequency band, i.e., same Gray code representation when quantized into binary bits while that of Eve is noticed to be different. 
\subsection{Quantization}
The power measurements are quantized into binary information bits using multi-level quantization technique with evenly spaced quantization levels. 
A Gray coding scheme is defined, where the number of bits per power measurement is given as $\log_2(Q)$ with $Q$ being the number of quantization levels. The obtained binary sequence at the end of this step is of length $r_l \in \{0,1\}^{K\log_2Q}$ for $K$ filters and $Q$ quantization levels.

\subsection{Information reconciliation} \label{sec:ir}
Due to noise or imperfect power estimation the observations at Alice and Bob could differ. Depending on the granularity of the quantization levels defined, this could lead to different binary sequences. To correct these errors one of the users (Alice) generates a syndrome $\mathbf{s}_A$ using distributed source coding techniques and this $\mathbf{s}_A$ is transmitted over a public channel. The second user (Bob) uses this information to correct errors through an error correction codes (ECC) decoders.  
The syndrome $\mathbf{s}_A$ transmitted over the public channel is also available to Eve. Using this syndrome and her initial measurements Eve might be able to decode part or all of the bit sequence or successfully reconcile. Thus, it is necessary to conservatively estimate potential information leakage in the privacy amplification step. 



\subsection{Privacy amplification}
In order to enable trustworthy communication, it must be ensured that the generated keys are secure and unknown to nearby eavesdroppers. 
In the current work, we assume that Eve measures the channel simultaneously as Bob, hence, information leakage depends on correlations with Eve's measurements and on the syndrome $s_A$ exchanged in the clear over the public channel. To account for the leaked information we evaluate the conditional min-entropy (CME)~\cite{cme}:  
\begin{equation}
    H_{\infty}(\mathbf{r}_A|\mathbf{r}_E, \mathbf{s}_A)=-\log_2\max_{\mathbf{r}_A \in \mathcal{R}_A, \mathbf{r}_E \in \mathcal{R}_E, \mathbf{s}_A \in \mathcal{S}_A}p(\mathbf{r}_A|\mathbf{r}_E, \mathbf{s}_A). \label{eq:cme}
\end{equation}
Here, $\mathcal{R}_A, \mathcal{R}_E, \mathcal{S}_A$ are the vector spaces for quantizer outputs of Alice and Eve and the syndrome shared by Alice respectively.
This metric estimates the number of secret bits in Alice's (Bob's) sequence when conditioned on Eve's channel measurements and the syndrome. 
As a result, the key size is upper bounded as
\begin{equation}
    |\mathbf{k}| \leq H_{\infty}(\mathbf{r}_A|\mathbf{r}_E, \mathbf{s}_A).  \label{eq:key_size}
\end{equation}
The evaluation of the conditional min-entropy in \eqref{eq:cme} requires knowledge on the underlying distributions of the variables. Obtaining a closed-form expression for a given environment is a complex task and is out of the scope of this paper. Instead, we use a conditional min-entropy numerical estimator, i.e., the fast black-box leakage estimation (F-BLEAU)~\cite{FBLEAU}. F-BLEAU evaluates the conditional min-entropy using machine learning based approaches, more precisely the nearest neighbor and frequentist estimates~\cite{leakiEst}. The estimator computes conditional min-entropy as the difference between min-entropy and leakage (this is in accordance to the work in~\cite{Quant_inf_flow}). Once the value is estimated the sequences at the legitimate users are compressed to a size in accordance to \eqref{eq:key_size}.

During our evaluation it was noticed that the min-entropy value, evaluated before conditioning, is consistent with another estimator which is part of the NIST recommended min-entropy test suite\footnote{The NIST min-entropy test suite consist of $10$ different tests. The tests evaluate only min-entropy and do not provide an estimate of leakage or conditional min-entropy. In this sense, the FBLEAU estimator has an advantage over these tests.}~\cite{NIST_min_entropy}, namely the most common values estimate. However, it has been reported that the NIST suite might under / over estimate the actual value~\cite{9393401}. As a measure of precaution, in the privacy amplification we compress $10\%$ more than the estimated value by F-BLEAU.   

\section{Measurement Campaign}\label{sec:measurement_campaign}

\begin{figure}[!t]
\centering
\includegraphics[clip, trim=6.5cm 0cm 4.3cm 2cm,width=0.48\textwidth]{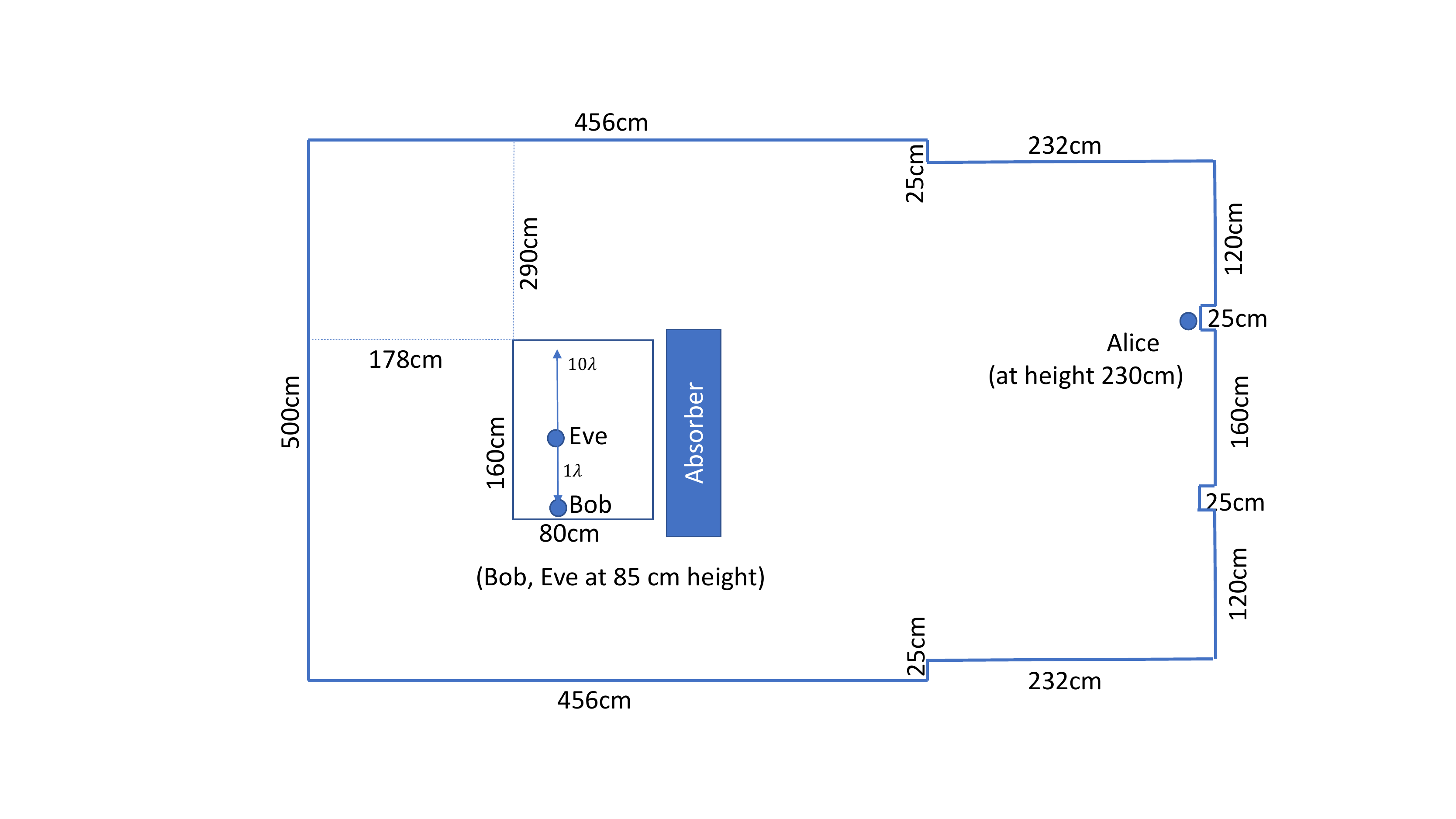}
\vspace{-1 cm}
\caption{The distance between Alice and Bob is $4$ m. Alice is placed at $2.3$ m height. Bob and Eve are at $0.85$ m height. Throughout the measurement campaign Eve is placed on a linear positioner which moves her with one-wavelength steps with respect to Bob's position (from $2\lambda$ to $6\lambda$).  }
\label{fig:lab_dimensions}
\vspace{-0.4cm}
\end{figure}

\begin{figure}[!t]
    \centering
    \includegraphics[width=0.50\textwidth]{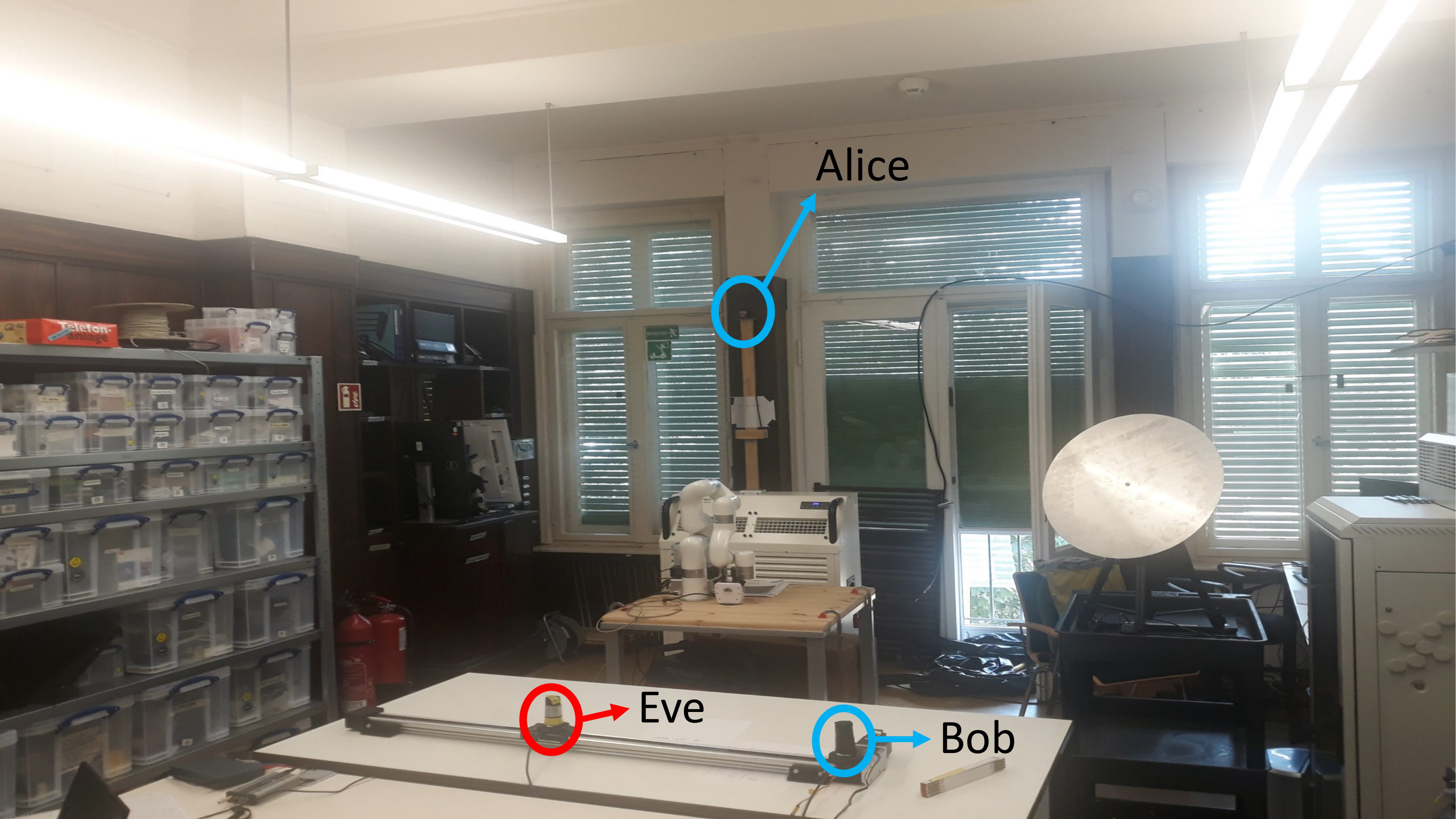}
    \caption{Line-of-sight scenario. Dynamic and static cases differentiate based on whether there is movement in the room. Eve is placed on a linear positioner and is moved from $2\lambda$ up to $6\lambda$ with respect to Bob's position. }
    \label{fig:los}
\end{figure}

\begin{figure}[!t]
    \centering
    \includegraphics[width=0.50\textwidth]{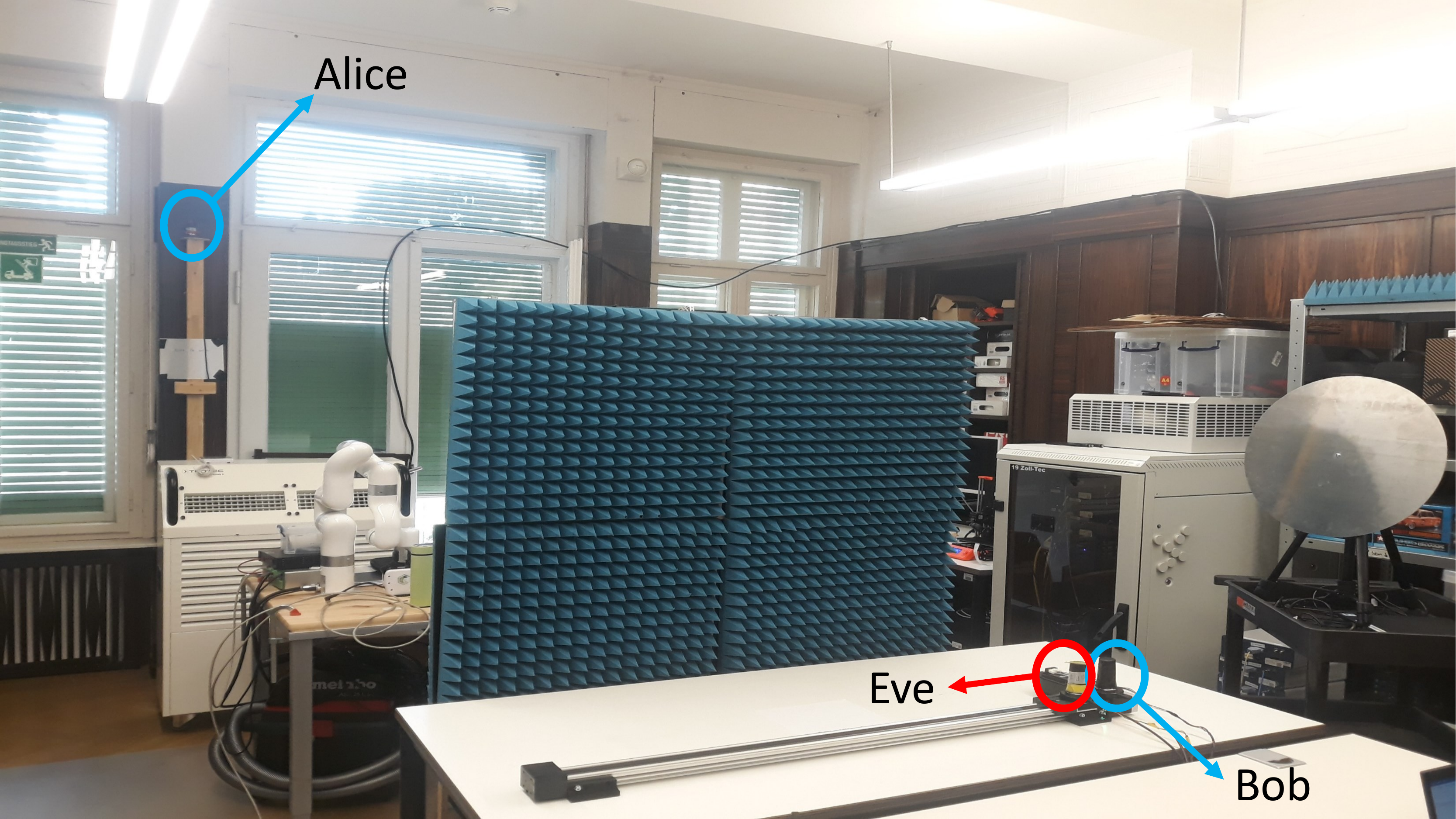}
    \caption{Non-line-of-sight scenario achieved using a set of absorbers to block the direct path between nodes. Dynamic and static cases differentiate based on whether there is movement  in the room. Eve is placed on a linear positioner and is moved from $2\lambda$ up to $6\lambda$ with respect to Bob's position. }
    \label{fig:nlos}
    \vspace{-0.4cm}
\end{figure}

For our experiment, we configured three universal software radio peripherals (USRPs), each with a single antenna, as two legitimate users, Alice and Bob and an eavesdropper, Eve. In more detail, USRP-2974 from National Instruments were used. Fig.~\ref{fig:lab_dimensions} shows the measurement setup. Experiments were performed in $4$ scenarios, namely line of sight (LoS) static, LoS dynamic, non line of sight (NLoS) static and NLoS dynamic. Dynamic scenarios were realized through movements of objects and people in the room.
Static channel measurements were performed during nighttime when there is no movement in the room. The LoS and NLoS scenarios were created by the absence or presence, respectively, of absorbers between the antennas of Alice and Bob. Figs~\ref{fig:los} and ~\ref{fig:nlos} show the LoS and NLoS scenarios. One of the objects moved throughout the measurement campaign during the dynamic channel measurements, i.e., a circular metal plate, is also displayed in the figure. 

Alice and Bob transmitted complex chirp signals in TDD. For the considered passband frequency, $f_c = 3.75$ GHz, the wavelength $\lambda\approx$ 8 cm.
Eve is placed at $5$ different positions with distances of $2\lambda, 3\lambda, 4\lambda,5\lambda,6\lambda$ w.r.t. Bob, i.e., at a distance of $16, 24, 32,$ etc. cm (on-the-shoulder attack). As a passive eavesdropper, she recorded all exchange between the legitimate users. 
In order to guarantee convergence of the statistical conditional min-entropy and leakage estimations, $10^5$ chirp signals were exchanged at each of the positions of Eve. The signal bandwidth was $B= 70$ MHz, the sampling rate was $f_s = 140$ MHz and the symbol duration was $T_s = 17.1875$ $\mu$s.

To implement the SKG protocol explained in Section~\ref{sec:SKG}, Alice, Bob and Eve convolve their received time domain measurements $y_A, y_B, y_E$ with a filterbank. A number of design parameters were implemented, i.e, different number of filters, different number of quantization levels, different code rates and decoder types. However, due to space constraints, in this work, we focus only on a subset of combinations. A comprehensive overview on our measurement campaign can be found in~\cite{jcas}. 

In this work, we consider a filterbank with $K = 16$
raised cosine filters with roll-off factor of $0.25$. Power measurements at the output of each filter are then converted into binary values using multi-level quantization with $Q=\{4, 16\}$ quantization levels. The quantization levels are evenly defined based on the range of the power measurements per channel realization. The information reconciliation is performed with Slepian-Wolf Polar codes. The syndrome $\mathbf{s}_A$, generated by Alice, is used by Bob and Eve to correct errors. The syndrome size depends on the code-rate $r = \{0.1,0.3,0.5,0.7,0.9\}$ and is given by $\mathbf{s}_A \in \{0,1\}^{(1-r)K\log_2Q}$. The success of the reconciliation depends on the syndrome, position and number of errors. The conditional min-entropy $H_{\infty}(\mathbf{r}_A|\mathbf{r}_E, \mathbf{s}_A)$ is evaluated to generate the shared secret key $\mathbf{k}$ between Alice and Bob.

\section{Results and Discussion}\label{sec:results}
In this section, we provide an overview of all SKG steps. 
\begin{figure}[!t]
    \centering
    \includegraphics[clip, trim=0cm 0cm 0cm 0cm,width=0.5\textwidth]{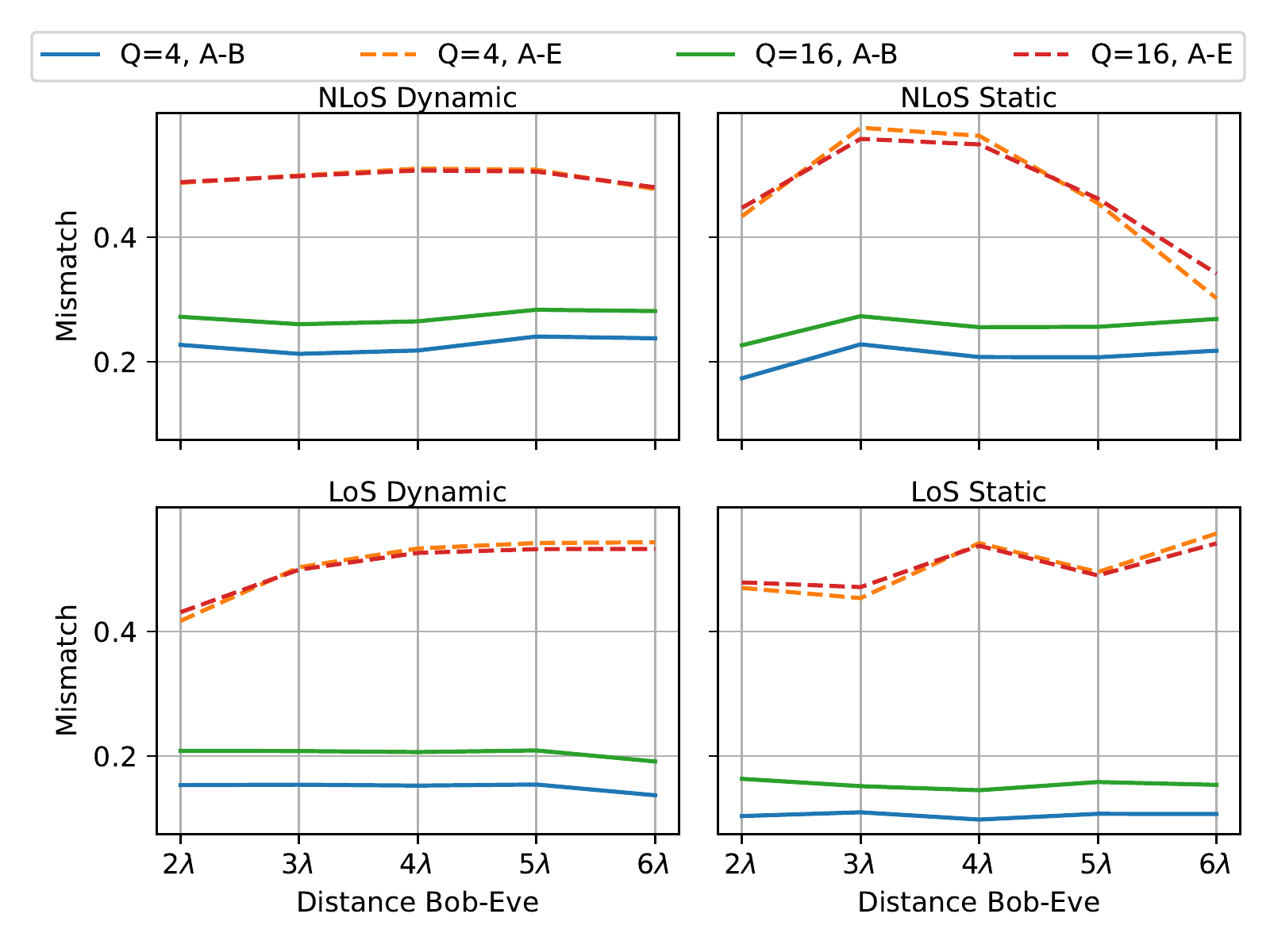}
    \caption{Mismatch between Alice-Bob (A-B) in solid and Alice-Eve (A-E) in dashed lines averaged over $10^5$ frames vs the distance between Bob and Eve, $1\lambda = 8$cm.}
    \label{fig:Mismatch}
\end{figure}
First, Fig.~\ref{fig:Mismatch} shows the mismatch probability between the binary bit sequences generated after quantization for $Q=\{4, 16\}$. The figure shows results for NLoS and LoS conditions for both static and dynamic environments.  
Notice that as Alice's and Bob's position do not change throughout the experiment, the mismatch probability remains stable. Different positions are considered for Eve and it can be seen that the mismatch probability at her end varies depending on the location. 

It is observed that the mismatch between Alice and Bob increases when the number of generated bits per sample increases (by increasing $Q$). This is due to  increased influence of non-reciprocal noise variations in the binary sequences. Mismatch is higher in NLoS as compared to LoS measurements. This could be attributed to a lower SNR as the dominant direct path is absent. 
On the other hand, the mismatch at Eve is not impacted from design parameters but from her location. Interestingly, it can be seen that the mismatch at Eve depends on the constructive and destructive interference of the multi-path components, i.e., on the surrounding environment rather than on absolute physical distance from Bob. 
For example in the NLoS static case, Eve has lower mismatch when she is located further from Bob, i.e., at distance of $3\lambda$ her mismatch is approximately $55\%$ while at $6\lambda$ distance the value can decrease to $30\%$.

\begin{figure}[!t]
    \centering
    \includegraphics[clip, trim=0cm 0cm 0cm 0cm,width=0.5\textwidth]{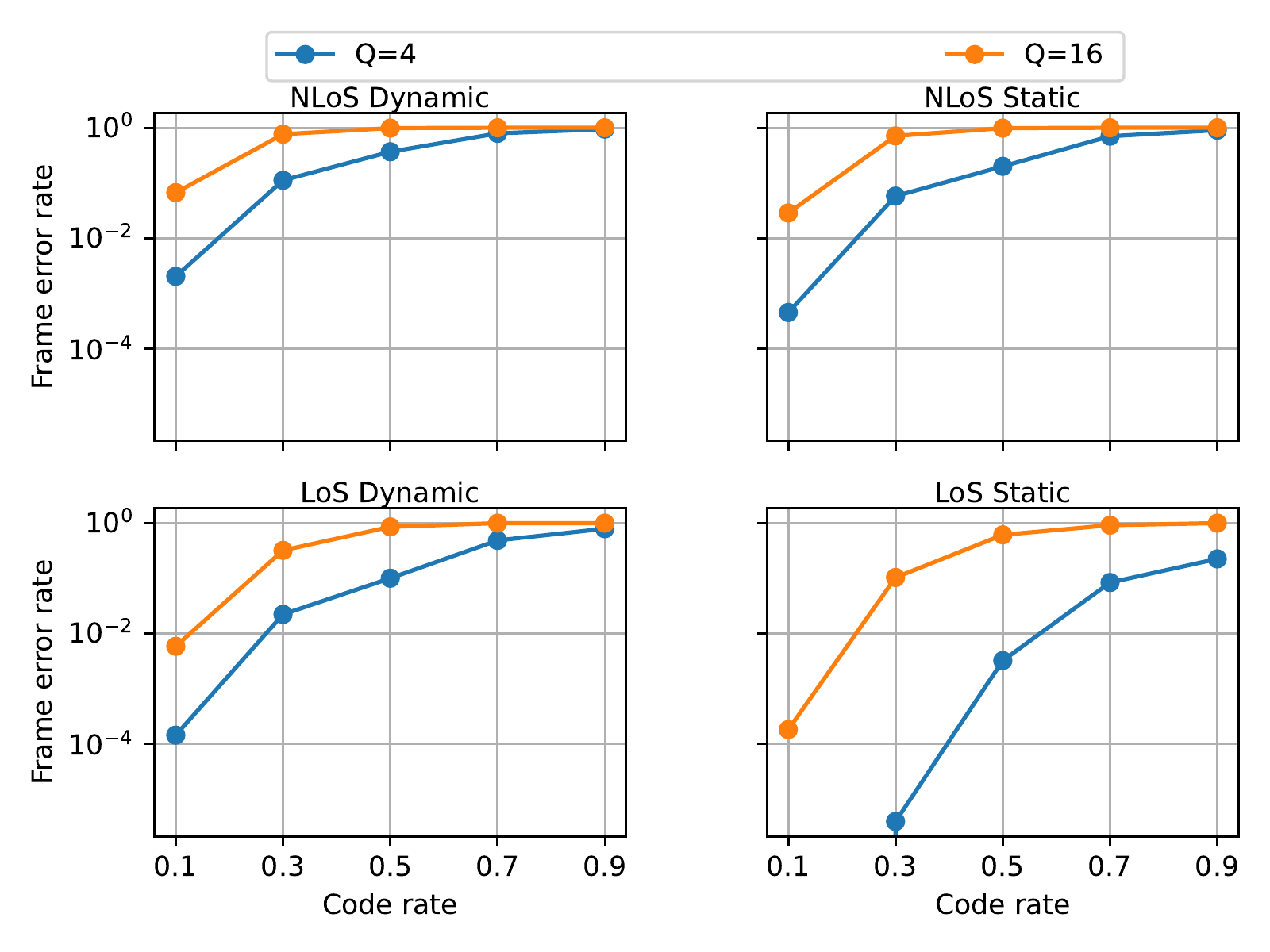}
    \caption{Frame error rates between Alice-Bob averaged over all 5 wavelengths.}
    \label{fig:FER_average}
\end{figure}
Next, Fig.~\ref{fig:FER_average}, depicts the frame error rate (FER) for Alice-Bob after reconciliation. We calculated the average FER across all locations of Eve since the mismatch between Alice and Bob is not affected by Eve's location. While we require lower code rates to achieve lower FERs, it is important to note that lower code-rates correspond to longer syndromes being sent over the public channel. We observe that the FER decreases if we consider lower number of quantization levels, $Q$, 
which aligns with the results in Fig.~\ref{fig:Mismatch}.

\begin{figure}[!t]
    \centering
    \begin{tikzpicture}
    \node[inner sep=0pt] at (0,4)
    {\includegraphics[clip, trim=0cm 0cm 1.5cm 0cm,width=0.5\textwidth]{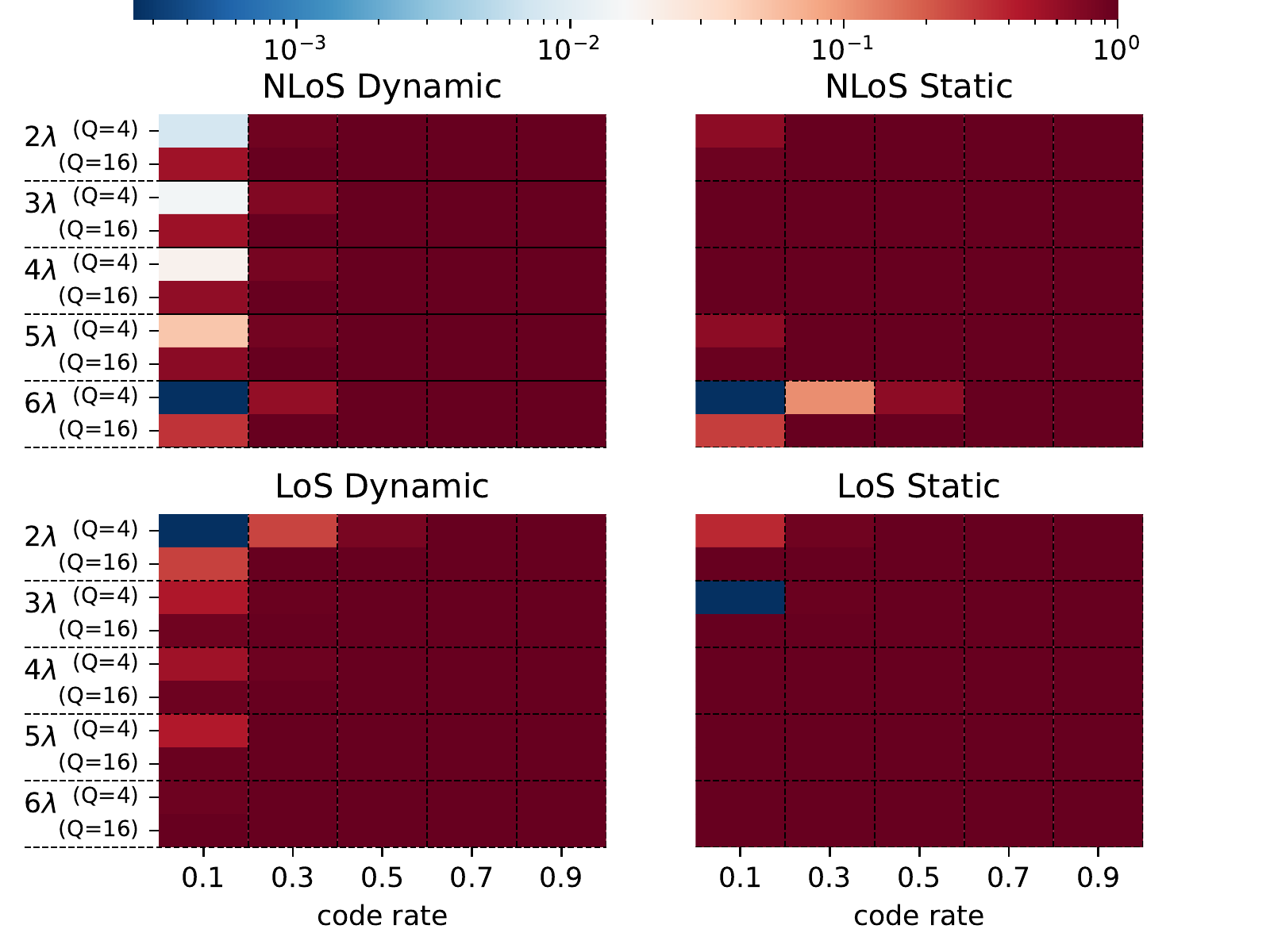}};
    \end{tikzpicture}
    \caption{Frame error rates between Alice-Eve for different wavelengths and quantization levels. }
    \label{fig:FER_Eve}
\end{figure}

Eve's FER is examined in Fig.~\ref{fig:FER_Eve}. As seen above, her position can play a critical role, therefore, we evaluate all locations. 
The FER for each location of Eve, as well as for each of $Q$ versus $r$, are presented as heat maps. This graph demonstrates that the presence of sufficient correlation, particularly when low values of $Q$ and $r$ are selected, increases Eve's likelihood of reconciling to the correct sequence. While red denotes $100\%$ FER it can be seen that some points on the map are different colors. It is observed that at some points (mostly at $r = 0.1$ and $Q = 4$) she can achieve FER less than $10^{-3}$. Hence, such parameters cannot be considered as a suitable for SKG in the given environment. Identifying potential leakage to Eve during the key generation process is crucial, therefore, it is essential to assess the suitability of the chosen design parameters and take into account the factors that may contribute to leakage.



\begin{figure}[!t]
    \centering
    \includegraphics[clip, trim=5.6cm 0cm 5.6cm 0cm,width=0.40\textwidth]{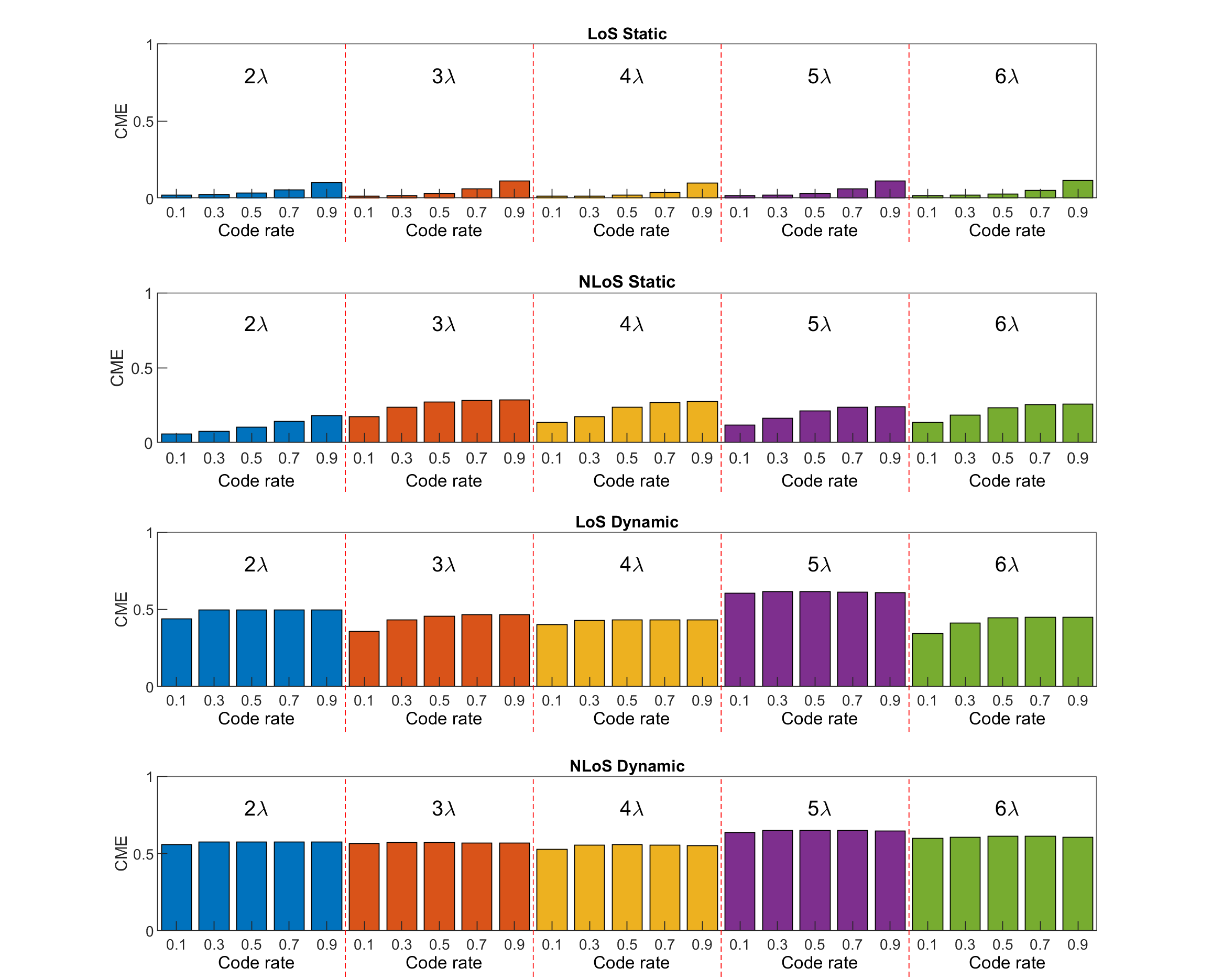}
    \caption{CME for all 5 wavelengths and code-rates for $Q=16$.}
    \label{fig:CME}
\end{figure}

\begin{figure}[!t]
    \centering
    \includegraphics[clip, trim=0cm 0cm 0cm 0cm,width=0.48\textwidth]{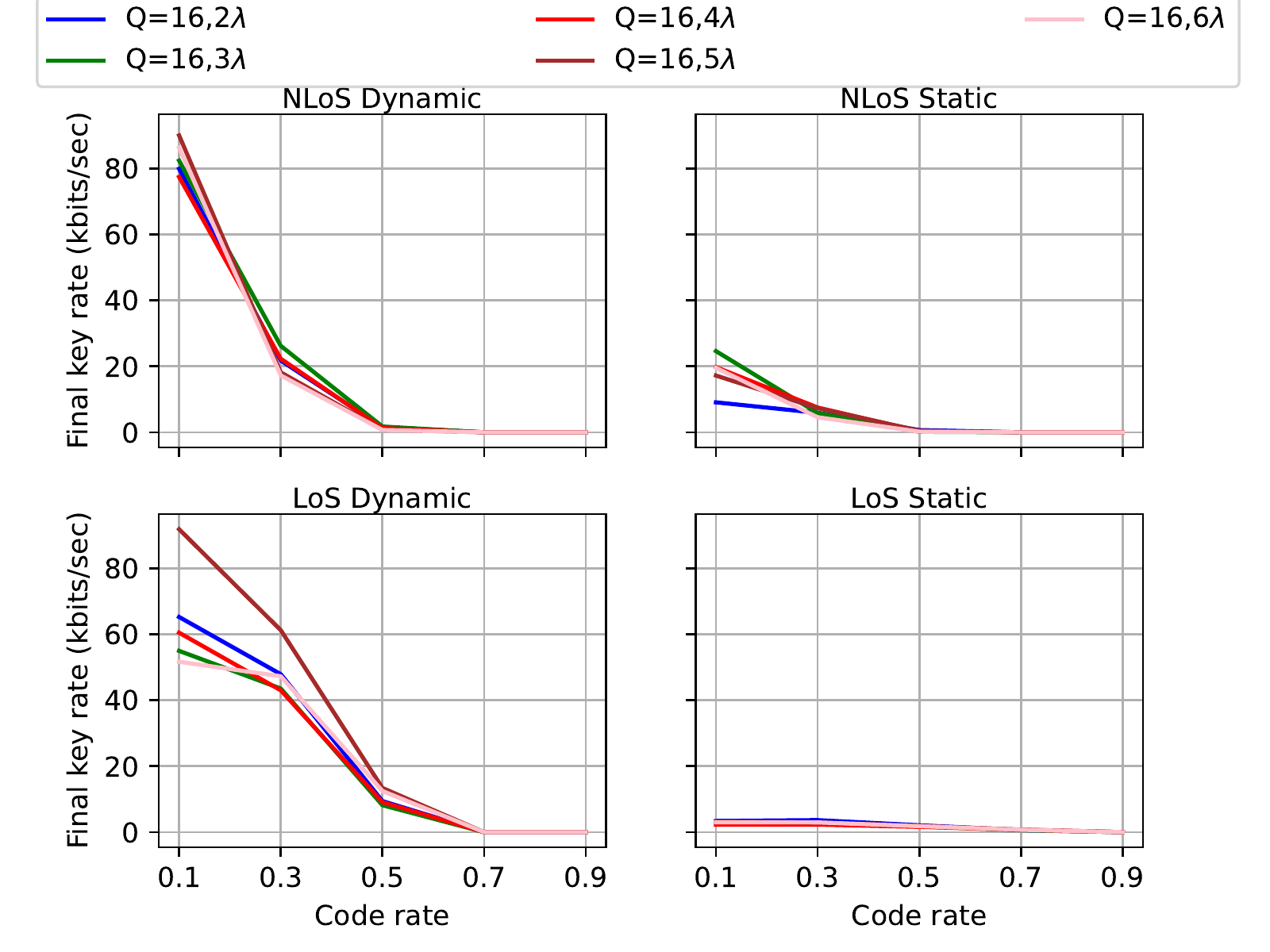}
    \caption{Final key rates between Alice-Bob in the  for all 5 wavelengths when $Q = 16$. }
    \label{fig:SKG_rate}
\end{figure}

Since $Q = 4$ is shown to allow Eve to retrieve the reconciled sequences, we continue our evaluation considering only $Q = 16$(as mentioned before, the full analysis of the design space is included in \cite{jcas}). As a next step we estimate the leakage to Eve. Fig.~\ref{fig:CME} represents the numerical evaluation of conditional min-entropy (CME) using the F-BLEAU estimator. We note that, in static channels this value is almost zero and we attribute this to the lack of channel entropy.
In case of dynamic channels, the randomness due to the multi-path channel variations results in higher CME. We also note that as we increase the code rate, the CME increases as the length of the leaked syndrome decreases. 
We conclude that we need to account for both the FERs and the CME. We define the final key rate achieved follows:
\begin{equation}
    R = K \times \log_2(Q) \times (1-FER) \times \hat{H}_{\infty}(\mathbf{r}_A|\mathbf{r}_E, \mathbf{s}_A),
\end{equation}
where, $\hat{H}_{\infty}(\mathbf{r}_A|\mathbf{r}_E, \mathbf{s}_A)$ is the numerical CME estimation from F-BLEAU increased by an additional $10\%$ to account for estimation inaccuracies. In Fig.~\ref{fig:SKG_rate}, we provide the final key rates for different combinations of code rates and measurement instances (i.e., different positions of Eve and channel measurement scenarios) when using $Q = 16$. Overall, it can be concluded that SKG rate is channel specific, i.e., it is highly dependent on both the environment and the setup.  

From \eqref{eq:key_size}, we infer that in order to generate 256 bit keys using SHA-256, the input sequence length, $input\_key$ needs to meet the upper bound,
\begin{math}
    |\mathbf{input\_key}| \leq 256/H_{\infty}(\mathbf{r}_A|\mathbf{r}_E, \mathbf{s}_A).  \label{eq:key_size_sha256}
\end{math}
From Fig.~\ref{fig:CME}, the minimum estimated CME is $0.014$ for LoS Static, $4\lambda$, $r = 0.1$ combination requiring 18, 285 bits of input sequence for the SHA-256 hash function. The maximum estimated CME is $0.615$ for NLoS Dynamic, $5\lambda$, $r = 0.3$ combination and hence we need 417 bits to generate a 256 bit hashed key .

\section{Security Challenge}\label{sec:sec_challenge}
With this paper we announce the security challenge in PLS. Using the approach presented in the previous sections, we generated a set of keys between Alice and Bob which are hashed so as to comply with \eqref{eq:key_size}. For the security challenge we consider the following design parameters, $K = 16$, $Q = 16$ and $r = 0.3$. The maximum CME for this combination is $0.615$ (NLoS Dynamic, $5\lambda$) and the minimum value of CME is $0.015$ (LoS Static, $4\lambda$). We thus require 417 and 17,0617 inputs bits for maximum and minimum combinations respectively to be hashed using SHA-256.
To arrive at the required input lengths, we concatenate successive successfully reconciled frames between Alice and Bob starting at different indices for each combination. Once the required length is achieved, we apply the SHA-256 hash function to obtain the final keys. 

$20$ of the legitimate keys are used to encrypt $20$ different blocks of plaintext (each of 256 bits) using one-time pad. In detail, we simply XOR each of the legitimate keys with a corresponding block of plaintext to create the encrypted ciphertexts. Specifically, one key is generated for each position of Eve ($2\lambda, 3\lambda, 4\lambda, 5\lambda, 6\lambda$) and each measurement scenario (NLoS Dynamic, NLoS Static, LoS Dynamic and LoS Static), resulting in a total of $20$ blocks of ciphertext, each encrypted using a 256-bit key derived for the corresponding measurement combination. The challenge aims at scrutinizing the SKG implementation, as opposed to an encryption algorithm, and it is exactly for this reason that we chose not to use the keys with quantum resistant symmetric block ciphers (e.g., AES-256 in GCM mode) but simply with a one time pad. We give all $10^5$ Eve's received signals in time domain for each of the positions presented in this paper. The syndromes produced by Alice, as described in Section~\ref{sec:ir}, are included, along with the starting frame index utilized for generating the input bit sequences for the SHA-256 hash function. Additionally, the respective conditional min-entropy is provided for all 20 combinations of Eve's position and measurement scenario.

We call all readers to attempt and regenerate the secret keys at Alice and Bob using the observations and public information at the eavesdropper. An attack
will be considered successful if part or all of the plaintext blocks are retrieved though processing (not random guessing). Channel measurements and syndromes are available at the IEEE dataport and can be accessed at \cite{security_challenge_BI}. Along with that, we provide detailed instructions and a Python script on how to load the data. In addition, we provide all the functions required for the SKG protocol, i.e., for the filterbank, the quantization and the reconciliation codes used. 

\section{Conclusions}\label{sec:conclusion}
This paper demonstrated that, in order to securely implement the SKG protocol, users need to be channel aware. The vulnerability of SKG to on-the-shoulder eavesdropping attacks is strongly dependent on the design parameters. The degree of the channel decorrelation between legitimate and adversarial users depends on the interference patterns of the multi-path observed at the eavesdropper rather than simply on the physical distance between the eavesdropper and the legitimate users. Overall, the success of the SKG protocol should rely on conservative measures and the conditional min-entropy should be carefully evaluated.

\bibliographystyle{IEEEtran}
\bibliography{Wiley}
\end{document}